\documentclass[12pt,preprint]{aastex}
\usepackage{graphicx}
\newcommand{\De}{\ensuremath{{\rm D}}}

\newcommand{\He}{\ensuremath{\null^4{\rm He}}}

\newcommand{\Li}{\ensuremath{\null^7{\rm Li}}}

\newcommand{\OmegaM}{\Omega_{m}}
\newcommand{\OmegaR}{\Omega_{r}}
\newcommand{\OmegaL}{\Omega_{\Lambda}}

\begin{document}

\title{{\bf Time variation of the electron mass in the early universe and the Barrow-Magueijo model}} 

\author{Claudia G. Sc\'{o}ccola \altaffilmark{1,2,a}} 
\email{cscoccola@fcaglp.unlp.edu.ar}
\and  
\author{Mercedes E. Mosquera \altaffilmark{1,a}} 
\email{mmosquera@fcaglp.unlp.edu.ar} 
\and 
\author{Susana J. Landau \altaffilmark{1,3,b}} 
\email{slandau@df.uba.ar} 
 \and 
\author{H\'ector Vucetich \altaffilmark{1,b}}
\email{vucetich@fcaglp.unlp.edu.ar}

\altaffiltext{1}{Facultad de Ciencias Astron\'{o}micas y
  Geof\'{\i}sicas. Universidad Nacional de La Plata. Paseo del 
 Bosque S/N 1900 La Plata, Argentina}
\altaffiltext{2}{Instituto de Astrof{\'\i}sica La Plata} 
\altaffiltext{3}{Departamento de F{\'\i}sica, FCEyN, Universidad de
  Buenos Aires, Ciudad Universitaria - Pab. 1, 1428 Buenos Aires,
  Argentina}

\altaffiltext{a}{fellow of CONICET}

\altaffiltext{b}{member of  the Carrera del Investigador Cient\'{\i}fico y Tecnol\'ogico, CONICET}


\begin{abstract}
We put limits on the time variation of the electron mass in the early
universe using observational primordial abundances of $\De$, 
$\He$ and $\Li$, recent data from the Cosmic Microwave Background and
the 2dFGRS power spectrum.  Furthermore, we use these constraints
together with other astronomical and geophysical bounds from the late
universe to test Barrow-Magueijo's model for the variation in
$m_e$.  From our analysis we obtain  $-0.615 < G\omega/c^4 < -0.045 $ (3$\sigma$
interval) in disagreement with the result obtained in the original
paper.
\end{abstract}

\section{Introduction}
\label{Intro}
Time variation of fundamental constants over cosmological time scales
is a prediction of theories that attempt to unify all fundamental
interactions like string derived field theories
\citep{Wu86,Maeda88,Barr88,DP94,DPV2002a,DPV2002b}, related
brane-world theories \citep{Youm2001a,Youm2001b,branes03a,branes03b},
and Kaluza-Klein theories \citep{Kaluza,Klein,Weinberg83,GT85,OW97}.
 In order to study the possible variation in the fine structure
constant or the electron mass, theoretical frameworks based on first
principles, were developed by different authors
\citep{Bekenstein82,Bekenstein2002,BSM02,BM05}.

The predicted time behaviour of the fundamental constants depends on
which version of the theories is considered. Thus, bounds obtained
from astronomical and geophysical data are an important tool to test
the validity of these theories. In a previous work \citep{Mosquera07},
we have analyzed the variation in the fine structure constant in the
context of Bekenstein model. In this paper, instead, we study the
variation in the electron mass ($m_e$) in the context of the
Barrow-Magueijo model \citep{BM05}. Note that $m_e$ is not a
fundamental constant in the same sense as the fine structure constant
is. Hence, it could be argued that constraints on the time variation of
the Higgs vacuum expectation value ($<v>$), rather than $m_e$, are
more relevant.  Moreover, the possibility of a time variation of the
vacuum expectation value of a field seems more plausible than the time
variation of a gauge coupling constant. However, in the context of the
Barrow-Magueijo model, the relevant fundamental constant is $m_e$ and
thus we will focus on its possible variation. The
joint variation in the fine structure constant and $<v>$ in the early universe will be
analyzed in a forthcoming paper.

Constraints on $m_e$ variation over cosmological
time scales are available from astronomical and local methods. The
latter ones include geophysical methods (analysis of natural
long-lived $\beta$ decayers in geological minerals and meteorites) and
laboratory measurements (comparisons of different transitions in
atomic clocks). The astronomical methods are based mainly in the
analysis of spectra from high redshift quasar absorption systems.
Bounds on the variation in $m_e$ in the early universe can be obtained
using data from the Cosmic Microwave Background (CMB) radiation and
from the abundances of light elements.  These bounds are not as
stringent as the mentioned above but they are important because they
refer to a different cosmological epoch.

In this paper, we perform a careful study of the time variation of
$m_e$ in the early universe. First, we use all available abundances of
$\De$, $\He$ and $\Li$, the latest data from the CMB and the 2dFGRS power spectrum to put
bounds on the variation in $m_e$ without assuming any theoretical
model. Afterward, we use these bounds and others from astronomical
and geophysical data, to test Barrow-Magueijo theory.

In section \ref{nucleo}, we use the abundances of the light elements
to put bounds on $\frac{\Delta m_e}{\left(m_e\right)_0}$, where
$\left(m_e\right)_0$ is the present value of $m_e$, allowing the
baryon to photon density $\eta_B$ to vary. In section \ref{cmb}, we
use the three year WMAP data, other CMB experiments and the power
spectrum of the final 2dFGRS to put bounds on the variation in $m_e$
during recombination, allowing also other cosmological parameters to
vary. In sections \ref{quasars}, \ref{geo} and \ref{clocks} we
describe the astronomical and local data from the late universe.  In
section \ref{modelo}, we describe the Barrow-Magueijo model, and
obtain solutions for the scalar field that drives the variation in
$m_e$, for the early and late universe.  In section \ref{resultados}
we show our results. Finally, in section \ref{resumen} we discuss the
results and summarize our conclusions.

\section{Bounds from BBN}
\label{nucleo}
Big Bang Nucleosynthesis (BBN) is one of the most important tools to study the early universe. The
baryon to photon ratio $\eta_B$ or equivalently
the baryon asymmetry $\eta_B \equiv \frac{n_B - n_{\bar{B}}}{n_\gamma}$ can be determined by comparison
between theoretical calculations and observations of the abundances of
light elements. An independent method for determining $\eta_B$ is
provided by data from the Cosmic Microwave Background (CMB)
\citep{wmapest,wmap3,Sanchez06}. Considering
the baryon density from WMAP results, the predicted abundances are
highly consistent with the observed $\De$ but not with all $\He$ and
$\Li$.  Such discrepancy is usually ascribed to non reported
systematic errors in the observations of $\He$ and $\Li$. However, if
the systematic errors of $\He$ and $\Li$ are correctly estimated, we
may have insight into new physics beyond the minimal BBN model.

In the currently most popular particle physics models, the lepton and
baryon numbers are comparable. In this case,  any asymmetry between
neutrinos and antineutrinos will not have a noticeable effect on the
predictions of BBN. However, observational data do not imply that the
lepton asymmetry should be connected to the `tiny' baryon asymmetry
$\eta_B$. Moreover, a small asymmetry between electron type 
neutrinos and antineutrinos can have a significant impact on BBN since the
$\nu_e$ affect the inter-conversion of neutrons to protons changing the
equilibrium neutron-to-proton ratio from $(n/p)^0_{eq}=e^{-\frac{\Delta
m}{T}}$ to $(n/p)_{eq}=(n/p)^0_{eq} e^{-\xi_e}$ where $\xi_e $ is the
ratio of the neutrino chemical potential to the temperature.
Consequently, the $\He$ abundance changes. In contrast, the $\De$
abundance is  insensitive to $\xi_e \neq 0$. Consistent with the BBN
and CMB data, values of $\xi_e$ in the range 
$-0.1 < \xi_e < 0.3$ are permitted
\citep{Barger03,Steigman05,Steigman06}. In this work, however, we
assume $\xi_e \simeq 0$ and attribute the discrepancies described
above to  time-variation of $m_e$ or $<v>$.

We considered available observational data on $\De$, $\He$ and
$\Li$. For $\De$, we used the values reported by
\citet{pettini,omeara,kirkman,burles1,burles2,Crighton04,omeara06,oliveira06}.
For $\He$, the available observations are reported by
\citet{PL07,izotov07}. 
For $\Li$, we considered the results from
\citet{ryan,bonifacio1,bonifacio2,bonifacio3,Asplund05,BNS05,bonifacio07}.
For the discussion about 
the consistency data check, we refer the reader to an earlier work
\citep{Mosquera07}.

We modified numerical code of Kawano \citep{Kawano88,Kawano92} in
order to allow $m_e$ to vary. The code was also updated with the
reaction rates reported by \citet{Iguri99}.  The main effects of the
possible variation in $m_e$ in the physics of the first three minutes
of the universe are changes in the weak rates, in the sum of electron
and positron energy densities, in the sum of electron and positron
pressures, and in the difference of electron and positron number
densities (see appendix \ref{correccion} for details). 
If $m_e$ takes a lower value than the present one, the primordial abundances are
higher than the standards. The change is more important for $\He$ and
$\Li$ abundances, where a variation of $10 \%$ in $m_e$
leads to a change of $7.4 \%$ and $8.5 \%$ in the abundances, while the
effect on the $\De$ abundance is tiny ($1.5 \%$).

\begin{figure}[h!]
\begin{center}
\includegraphics[scale=0.57,angle=0]{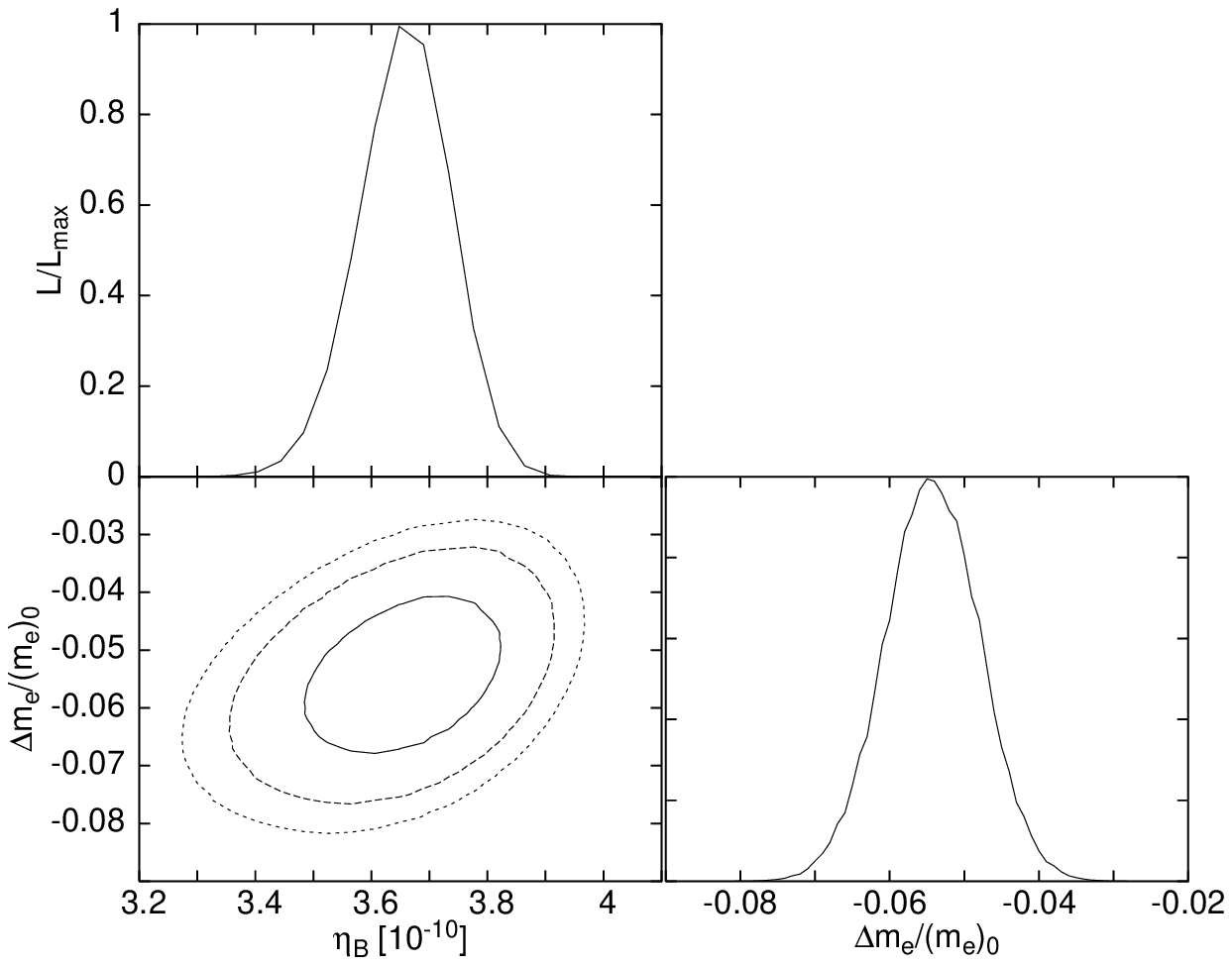}
\includegraphics[scale=0.57,angle=0]{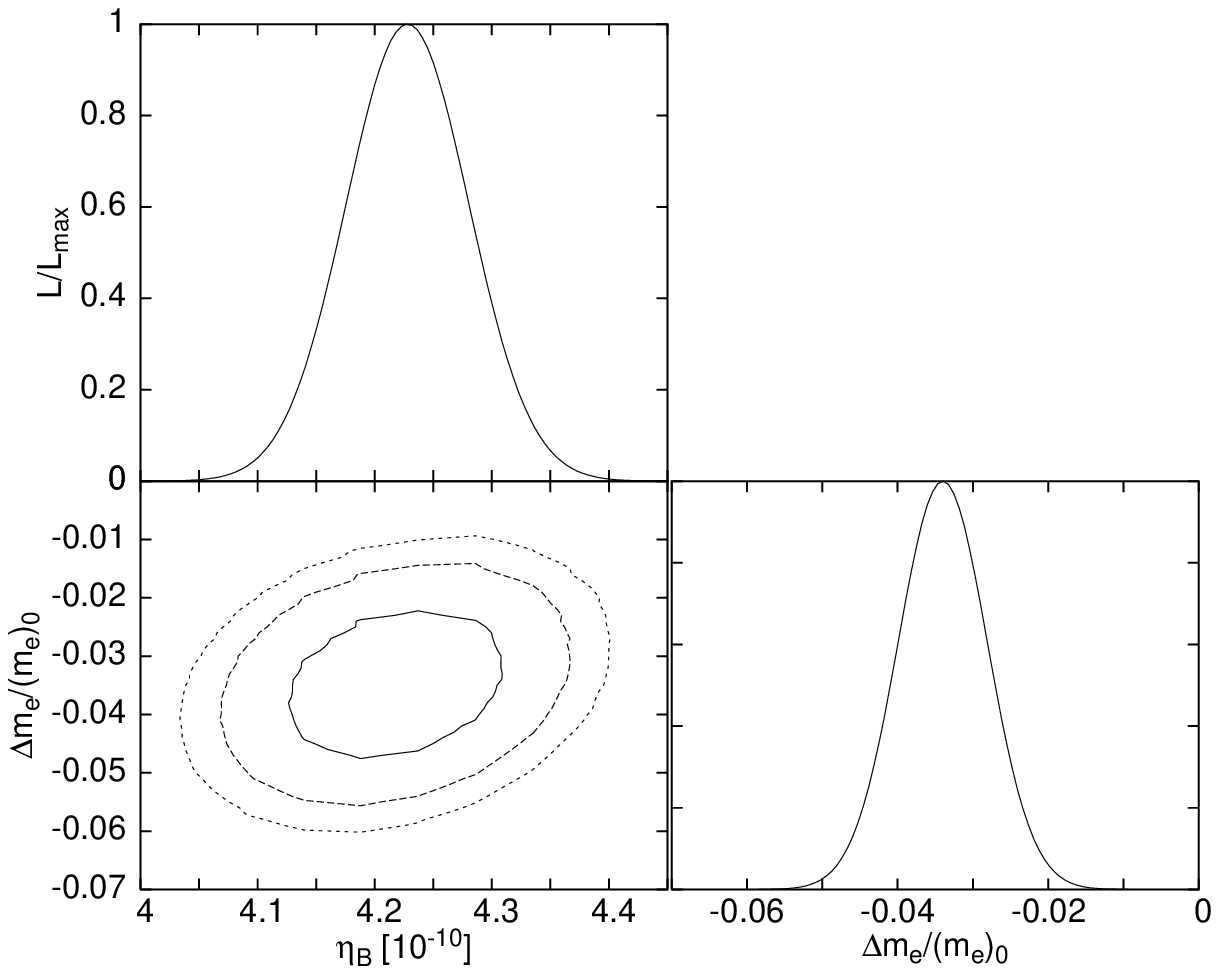}
\end{center}
\caption{$1 \sigma$, $2 \sigma$ and $3 \sigma$ likelihood contours for
$\frac{\Delta m_e}{\left(m_e\right)_0}$ vs $\eta_B$ and 1 dimensional Likelihood 
using $\He +\Li$ data (left) and all data (right) }
\label{contnucleo2}
\end{figure}

\begin{figure}[h!]
\begin{center}
\includegraphics[scale=0.57,angle=0]{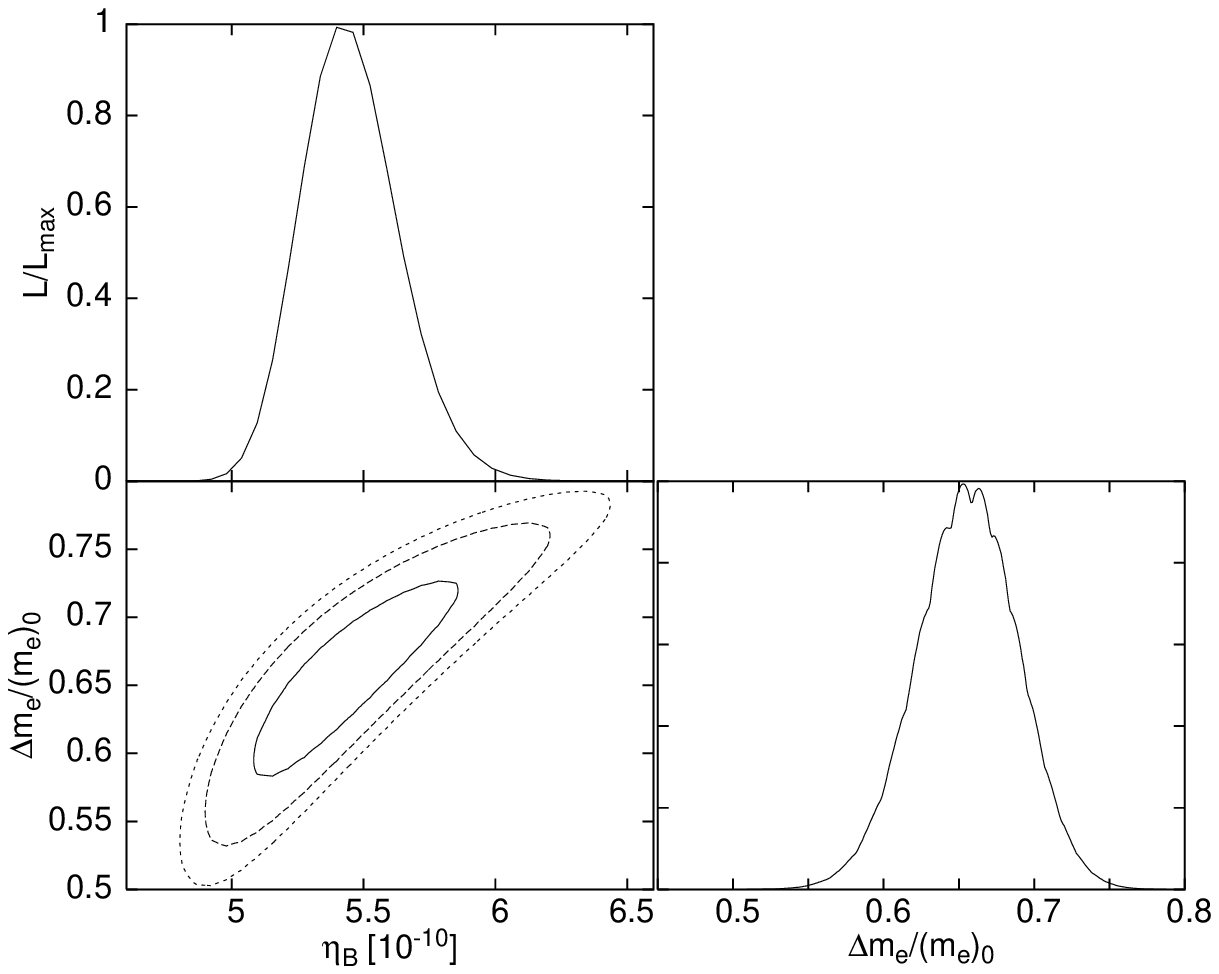}
\includegraphics[scale=0.57,angle=0]{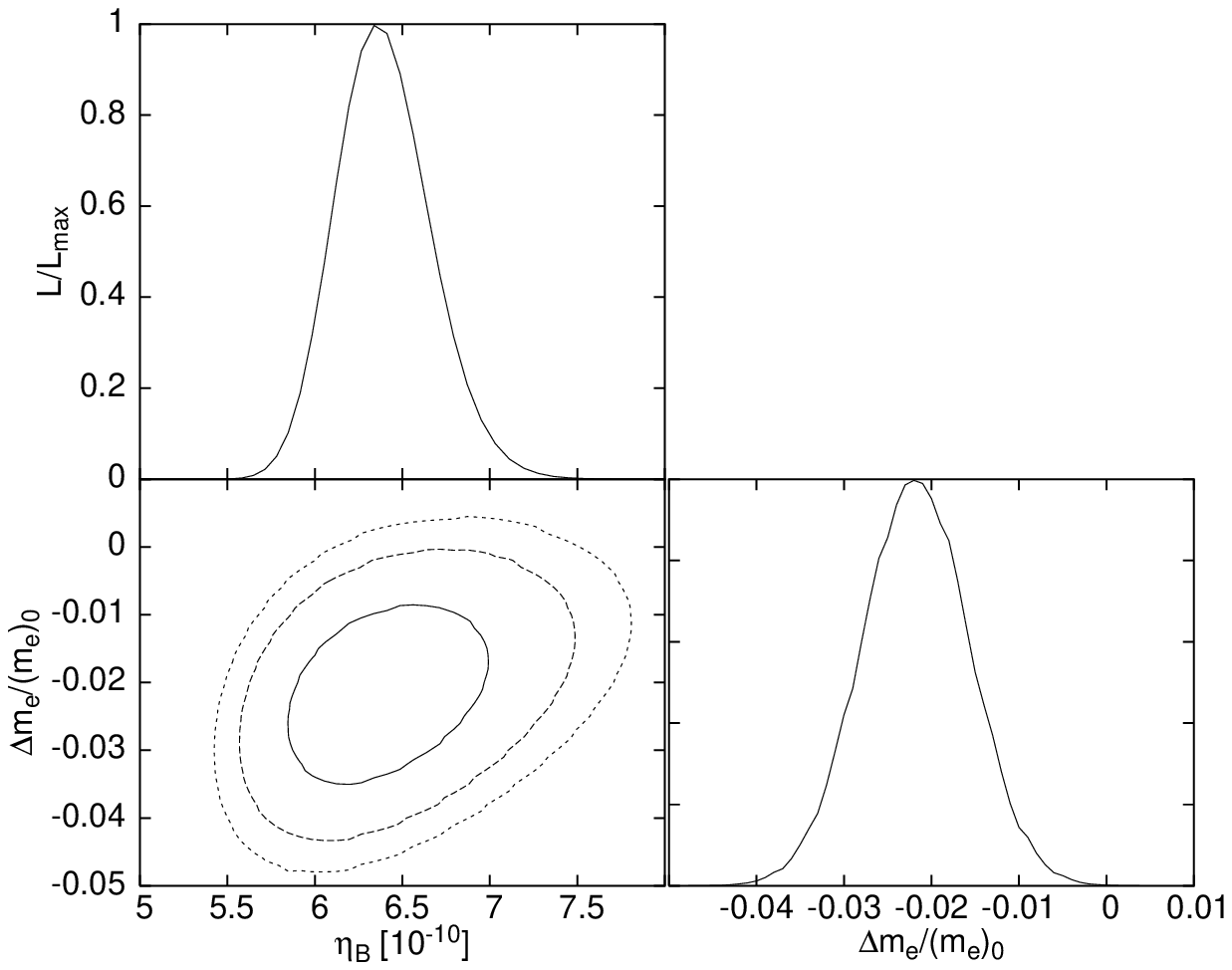}
\end{center}
\caption{$1 \sigma$, $2 \sigma$ and $3 \sigma$ likelihood contours for
$\frac{\Delta m_e}{\left(m_e\right)_0}$ vs $\eta_B$ and 1 dimensional
  Likelihood using $\De  + \Li$ data (left) and $ \De + \He $ (right)}
\label{contnucleo3}
\end{figure}

We computed the light nuclei abundances for different values
of $\eta_B$ and $\frac{\Delta m_e}{\left(m_e\right)_0}$ and performed
the statistical analysis to obtain the best fit values for these
parameters.
There is no good fit for the whole data
set even for $\frac{\Delta m_e}{\left(m_e\right)_0} \neq 0$.  However,
reasonable fits can be found excluding one group of data at each time
(see table \ref{me-tabla}). Figures \ref{contnucleo2} and
\ref{contnucleo3} show the confidence contours
and 1 dimensional Likelihoods for different groups of data.
\begin{table}[h!]
\begin{center}
\caption{Best fit parameter values and $1 \sigma$ errors for the BBN constraints on 
$\frac{\Delta m_e}{\left(m_e\right)_0}$ and $\eta_B$ (in units of $10^{-10}$).}
\label{me-tabla}
\begin{tabular}{|c|c|c|c|}
\hline
& $\eta_B\pm \sigma $&$\frac{\Delta m_e}{\left(m_e\right)_0}\pm
\sigma$&$\frac{\chi^2_{min}}{N-2}$ \\ \hline 
$\De + \He + \Li$ &  $4.237_{-0.097}^{+0.047}$&  $-0.036_{-0.007}^{+0.010}$ & 9.33
\\ \hline
$ \He +\Li $ &  $3.648_{-0.124}^{+0.128}$& $-0.055_{-0.008}^{+0.010}$& 1.00
 \\ \hline
 $\De + \Li$ &  $5.399_{-0.213}^{+0.287}$& $0.653_{-0.045}^{+0.051}$&1.01   \\
 \hline
 $\De + \He $ &  $6.339_{-0.355}^{+0.376}$ &
$-0.022 \pm 0.009$& 1.01 \\ \hline
\end{tabular}
\end{center}
\end{table}
We obtained that for $\De + \He$ the value
of $\eta_B$ is coincident with WMAP estimation and there is no
variation in $m_e$ within $3 \sigma$. Moreover, the other groups of data prefer values far from WMAP
estimation, and for $\De + \Li$, the result is consistent with
variation in $m_e$ within $6 \sigma$. 

As pointed out in the introduction, in the standard model, $\De$,
$\He$ and $\Li$ abundances considered separately predict very different values for the baryon
density. Therefore, when the three abundances are fitted together, an
intermediate value of $\eta_B$ is obtained, but the value of $\chi^2$
is too high. Only when two abundances are considered, we obtain a
reasonable fit. Furthermore, a high variation in $m_e$
which affects mostly the $\Li$ abundance is needed to fit $\De$ and
$\Li$ together. On the other hand, $\De$ and $\He$ are marginally
consistent with WMAP estimation and therefore no variation in $m_e$ is
needed to fit both data at the same time. Finally, in order to fit the
abundances of $\He$ and $\Li$, a variation in $m_e$  is needed since
both quantities are affected when $m_e$ is allowed to vary. 
%
%

As mentioned in the introduction, in this paper we limit ourselves to
the context of the Barrow-Magueijo model of a varying $m_e$. However,
in more general classes of theories (Kaluza-Klein, 
Strings, GUTs, etc), the underlying fundamental constant is the Higgs
vacuum expectation value. The dependence of the primordial abundances
on the Higgs vacuum expectation value has been analyzed by
\citet{YS03}.  Semi-analytical analysis have been performed by some of
us in earlier works \citep{Chamoun07}. Besides changes in $m_e$, a
possible variation in $<v>$ modifies the values of the 
following quantities: the Fermi constant $G_F$, the neutron-proton
mass difference $\Delta m_{np}$, and the deuterium binding energy
$\epsilon_D$ . The dependence of these quantities with $<v>$ have been
described in an earlier work \citep{Chamoun07} (see appendix
\ref{correccion} for details). We modified numerical code of Kawano in
order to allow $<v>$ to vary. 
The abundances of the primordial elements are much higher than the
standard value if the Higgs vacuum expectation value during BBN is
larger than the current value. 
A variation of $10 \%$ in $<v>$ leads to a change of $45
\%$, $25 \%$ and $29 \%$ in the $\He$, $\Li$ and $\De$ abundances
respectively. Since $\De$ is a residual of $\He$ production, a great
change in $\He$ also leads to an important change in $\De$. The
changes in the abundances are greater than in the case where only $m_e$ is allowed to vary.

\begin{figure}[h!]
\begin{center}
\includegraphics[scale=0.57,angle=0]{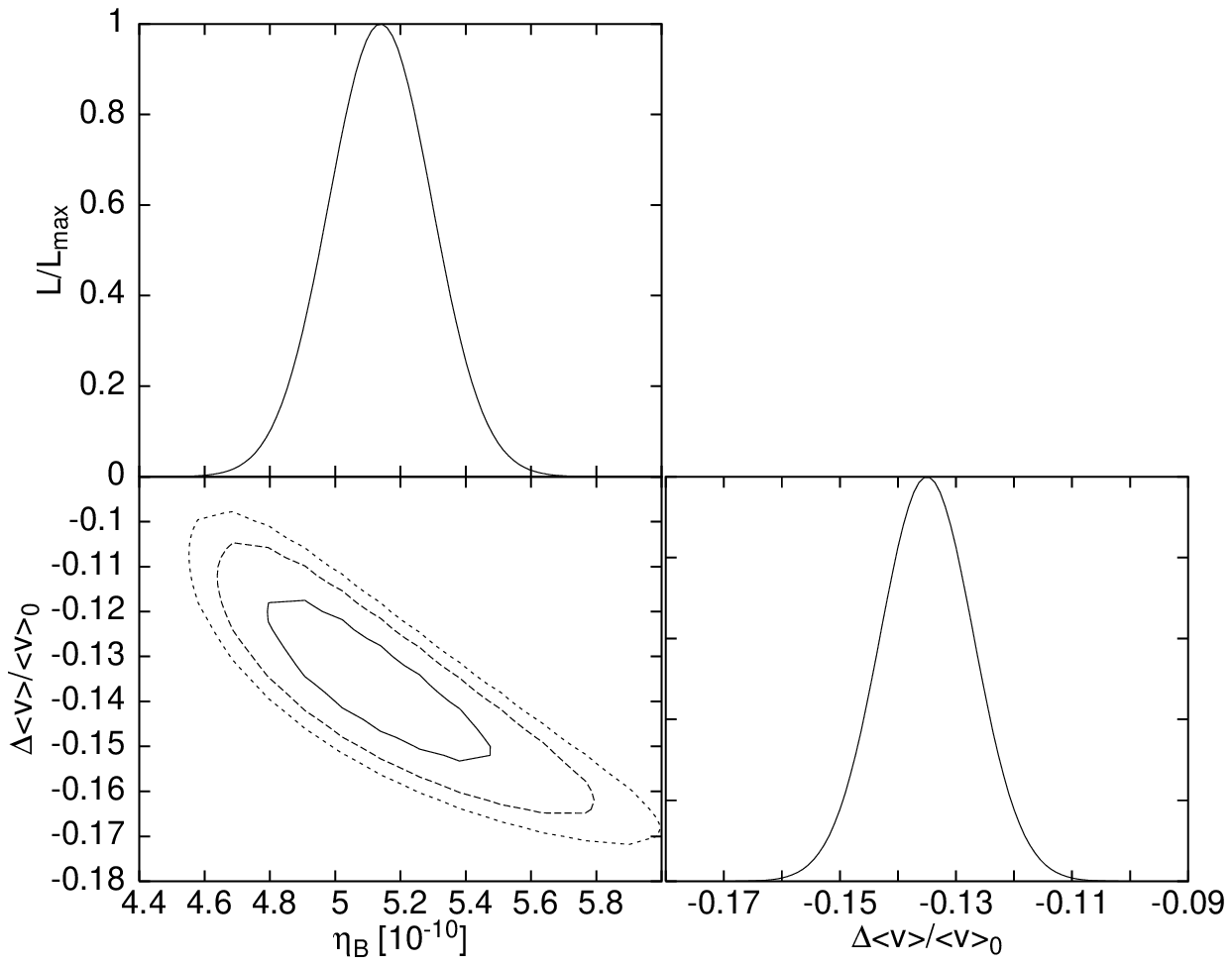}
\includegraphics[scale=0.57,angle=0]{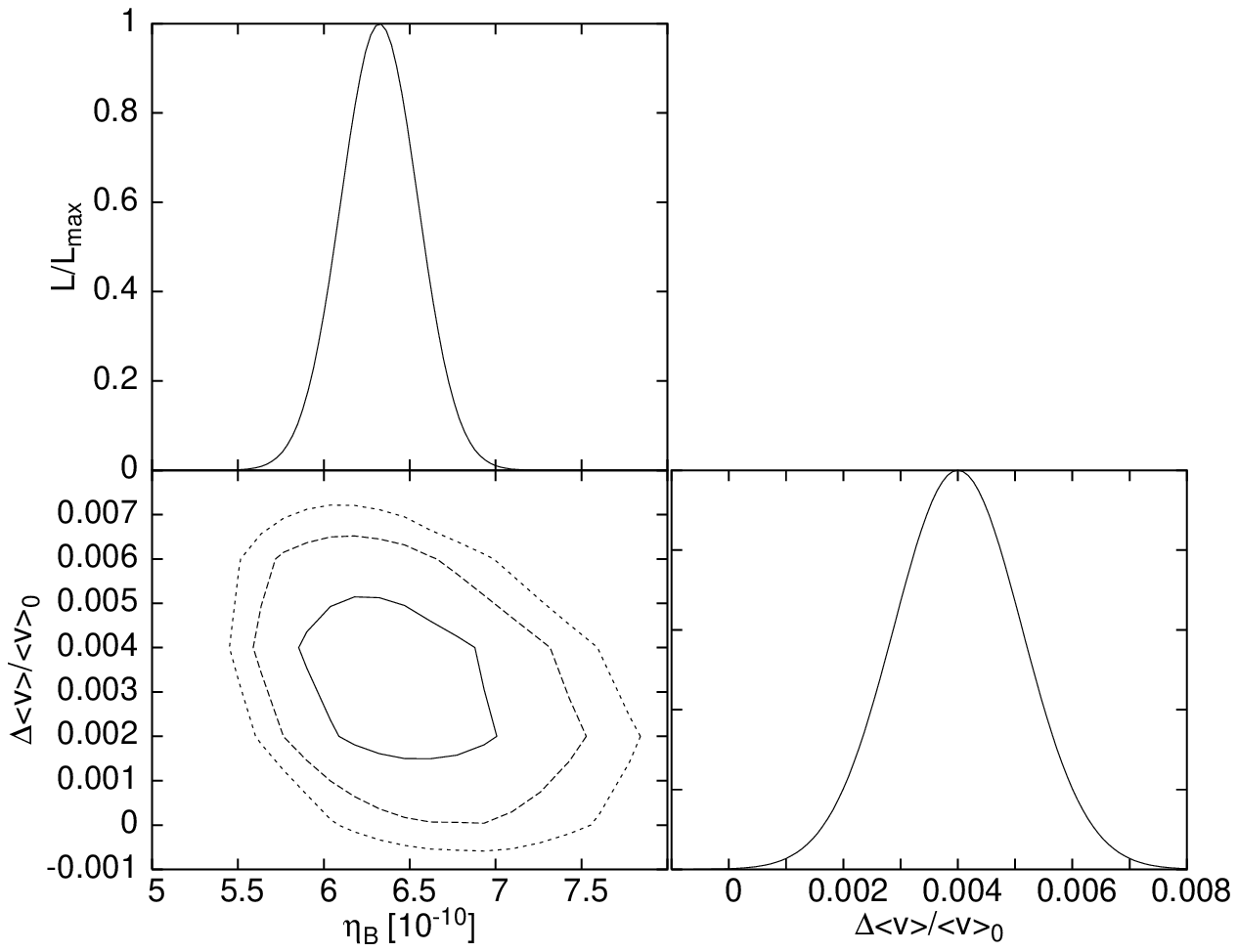}
\end{center}
\caption{$1 \sigma$, $2 \sigma$ and $3 \sigma$ likelihood contours for
$\frac{\Delta <v>}{<v>_0}$ vs $\eta_B$ and 1 dimensional
  Likelihood using $\De + \Li$ (left) and $\De + \He$ (right)}
\label{contnucleo4}
\end{figure}

In the case of $<v>$ , we performed the same analysis for the same
groups of data we considered for $m_e$. As in the case of $m_e$,
variation there is no good fit for the whole set of data. However,
reasonable fits can be found excluding one group of data at each time
(see table \ref{v-table}). Figure \ref{contnucleo4} shows the
confidence contours and 1 dimensional Likelihoods for different groups
of data.
\begin{table}[h!]
\begin{center}
\caption{Best fit parameter values and $1 \sigma$ errors for the BBN constraints on 
$\frac{\Delta <v>}{<v>_0}$ and $\eta_B$ (in units of $10^{-10}$).}
\label{v-table}
\begin{tabular}{|c|c|c|c|}
\hline
& $\eta_B\pm \sigma$&$\frac{\Delta <v>}{<v>_0}\pm \sigma$&$\frac{\chi^2_{min}}{N-2}$ \\ \hline
 $\De +  \He + \Li$&  $4.275 \pm 0.097 $& $0.006\pm 0.002$ & 9.27 \\ \hline
 $\He+\Li$ &  $3.723^{+0.132}_{-0.124}$&  $0.008\pm 0.001$& 1.00 \\ \hline
 $\De+\Li$ &  $5.139_{-0.231}^{+0.242} $& $-0.138_{-0.009}^{+0.015}$&  1.01 \\ \hline
 $\De+\He$&  $6.324_{-0.285}^{+0.374}$ & $0.004 \pm 0.002$& 1.04 \\ \hline
\end{tabular}
\end{center}
\end{table}

We obtain that for $\De + \He$ the value of $\eta_B$ is consistent
with WMAP estimation and there is no 
variation in $<v>$ within $3 \sigma$. Moreover, the other groups of
data prefer values not consistent with WMAP results. For $\De + \Li$,
the result is consistent with variation in $<v>$ within $6 \sigma$.  
The results are similar to those obtained in the case where $m_e$ is
the varying constant: 
i) no reasonable fit for the three
abundances; ii) $\De$ and $\He$ can be well fitted with null $<v>$
variation; iii) $\De$ and $\Li$ need a huge variation in order to
obtain a reasonable fit. However, the bounds on variation in $<v>$ are
more stringent  than the bounds obtained when only $m_e$ was
allowed to vary (see table \ref{me-tabla}). This could be explained
since variations in $<v>$ lead to greater changes in the theoretical
abundances than variation in $m_e$. 

\section{Bounds from CMB}
\label{cmb}
Cosmic Microwave Background (CMB) radiation provides valuable
information about the physical conditions of the universe just before
decoupling of matter and radiation, and thanks to its dependence upon
cosmological parameters, it allows their estimation.  Any change in
the value of $m_e$ affects the physics during
recombination, mainly the redshift of this epoch, due to a shift in
the energy levels and in particular, the binding energy of hydrogen.
The Thompson scattering cross section, which is proportional to
$m_e^{-2}$, is also changed for all particles. Therefore, the CMB
power spectrum is modified by a change in the relative amplitudes of
the Doppler peaks, and shifts in their positions (see appendix
\ref{apendice_recombinacion} for details).  Changes in the
cosmological parameters produce similar effects. In
the recombination scenario, the only effect of varying $<v>$ is a
change in the value of $m_e$.  Previous analysis of the
CMB data including a possible variation in $m_e$ have been performed by
\citet{YS03,KS00}. In this paper, we use the WMAP 3-year temperature
and temperature-polarization power spectrum \citep{wmap3}, and other
CMB experiments such as CBI \citep{CBI04}, ACBAR \citep{ACBAR02}, and
BOOMERANG \citep{BOOM05_polar,BOOM05_temp}, and the power spectrum of
the 2dFGRS \citep{2dF05}. We consider a spatially-flat cosmological
model with adiabatic density fluctuations. The parameters of our model
are:
\begin{equation}
P=(\Omega_B h^2, \Omega_{CDM} h^2, \Theta, \tau_{re}, \frac{\Delta m_e}{\left(m_e\right)_0}, n_s, A_s)
\end{equation}
where $\Omega_{CDM} h^2$ is the dark matter density in units of the
critical density, $\Theta$ gives the ratio of the comoving sound
horizon at decoupling to the angular diameter distance to the surface
of last scattering, $\tau_{re}$ is the reionization optical depth, $n_s$
the scalar spectral index and $A_s$ is the amplitude of the density
fluctuations.

\begin{figure}[h!]
\begin{center}\includegraphics[scale=0.7,angle=-90]{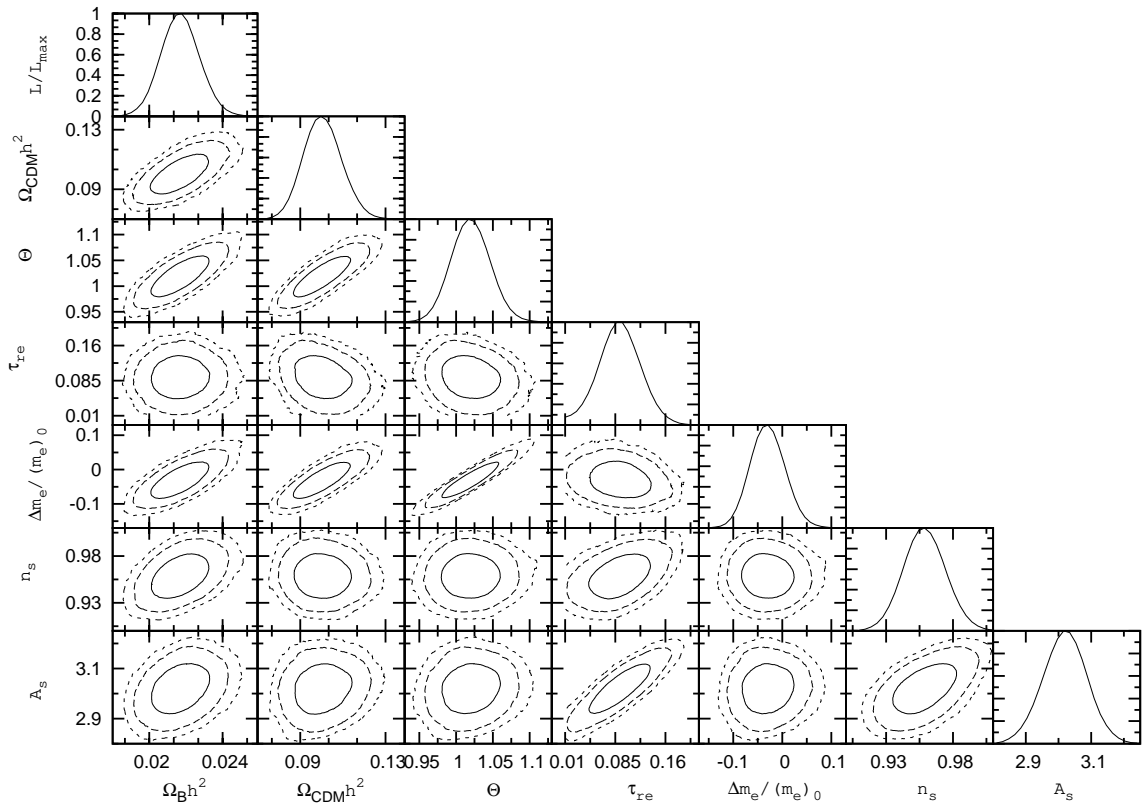}
\end{center}
\caption{Marginalized posterior distributions obtained with CMB data,
  including the WMAP 3-year data release plus 2dFGRS power
  spectrum. The diagonal shows the posterior distributions for
  individual parameters, the other panels shows the 2D contours for
  pairs of parameters, marginalizing over the others.}
\label{resulcmb}
\end{figure}

We use a Markov Chain Monte Carlo method to explore the parameter
space because the exploration of a multidimensional parameter space
with a grid of points is computationally prohibitive.  We use the
public available CosmoMC code of \citet{LB02} which uses CAMB
\citep{LCL00} and RECFAST \citep{recfast} to compute the CMB power
spectra, and we have modified them in order to include the possible
variation in $m_e$ at recombination. We ran eight different chains. We
used the convergence criterion of \citet{Raftery&Lewis} to stop the
chains when $R-1 < 0.0044$ (more stringent than the usually adopted
value).  Results are shown in table \ref{tablacmb} and figure
\ref{resulcmb}.  Figure \ref{resulcmb} shows a strong degeneracy
between $m_e$ and $\Theta$, which is directly related to $H_0$, and
also between $m_e$ and $\Omega_B h^2$, and $m_e$ and $\Omega_{CDM}
h^2$.
\begin{table}
\begin{center}
\renewcommand{\arraystretch}{1.3}
\caption{Mean values and errors for the main and derived parameters including $m_e$ variation 
($H_0$ is in units of $\rm km \, \, s^{-1} \, \,  Mpc^{-1} $).}
\label{tablacmb}
\begin{tabular}{|c|c|c|c|c|}
\cline{1-2} \cline{4-5}
  Parameter & Mean value and $1\sigma$ error &&Parameter & Mean value and $1\sigma$ error\\
\cline{1-2} \cline{4-5}
$\Omega_B h^2$ &  $0.0217 \pm 0.0010$&&
$\Omega_{CDM} h^2$ & $ 0.1006_{-0.0086}^{+0.0085}$ 
\\
\cline{1-2} \cline{4-5}
$\Theta$ & $1.020\pm 0.025$&&
$\tau_{re}$ & $0.091_{-0.014}^{+0.013}$ 
\\
\cline{1-2} \cline{4-5}
$\frac{\Delta m_e}{\left(m_e\right)_0}$ &  $-0.029\pm 0.034$&&
$n_s$ & $0.960\pm 0.015$
\\
\cline{1-2} \cline{4-5}
$A_s$ & $3.020\pm 0.064$&&
$H_0$ & $ 68.1_{-  6.0}^{+  5.9}$ 
\\
\cline{1-2}\cline{4-5}
\end{tabular}
\end{center}
\end{table}

\begin{figure}[h!]
\begin{center}
\includegraphics[scale=0.45,angle=-90]{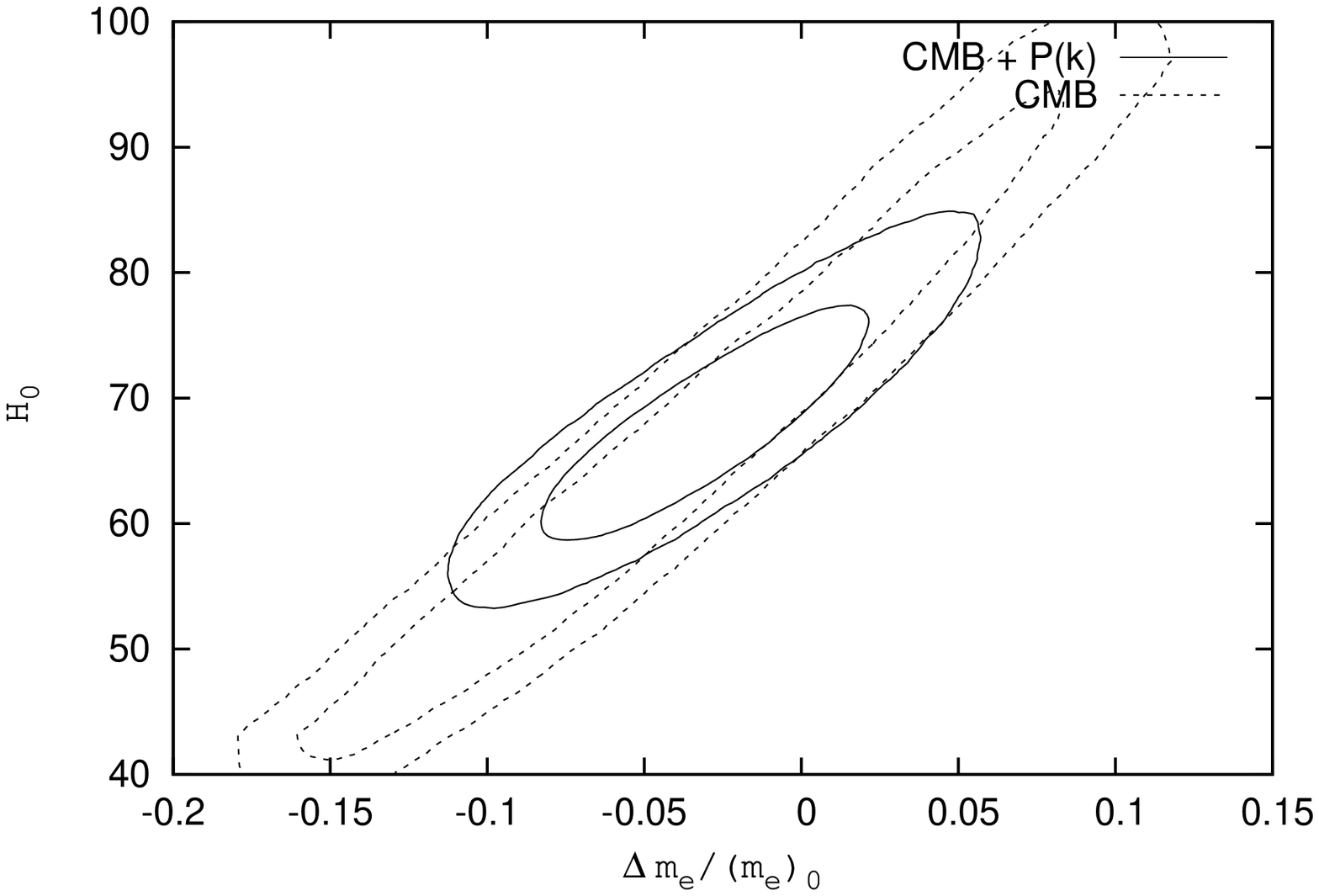}
\includegraphics[scale=0.45,angle=-90]{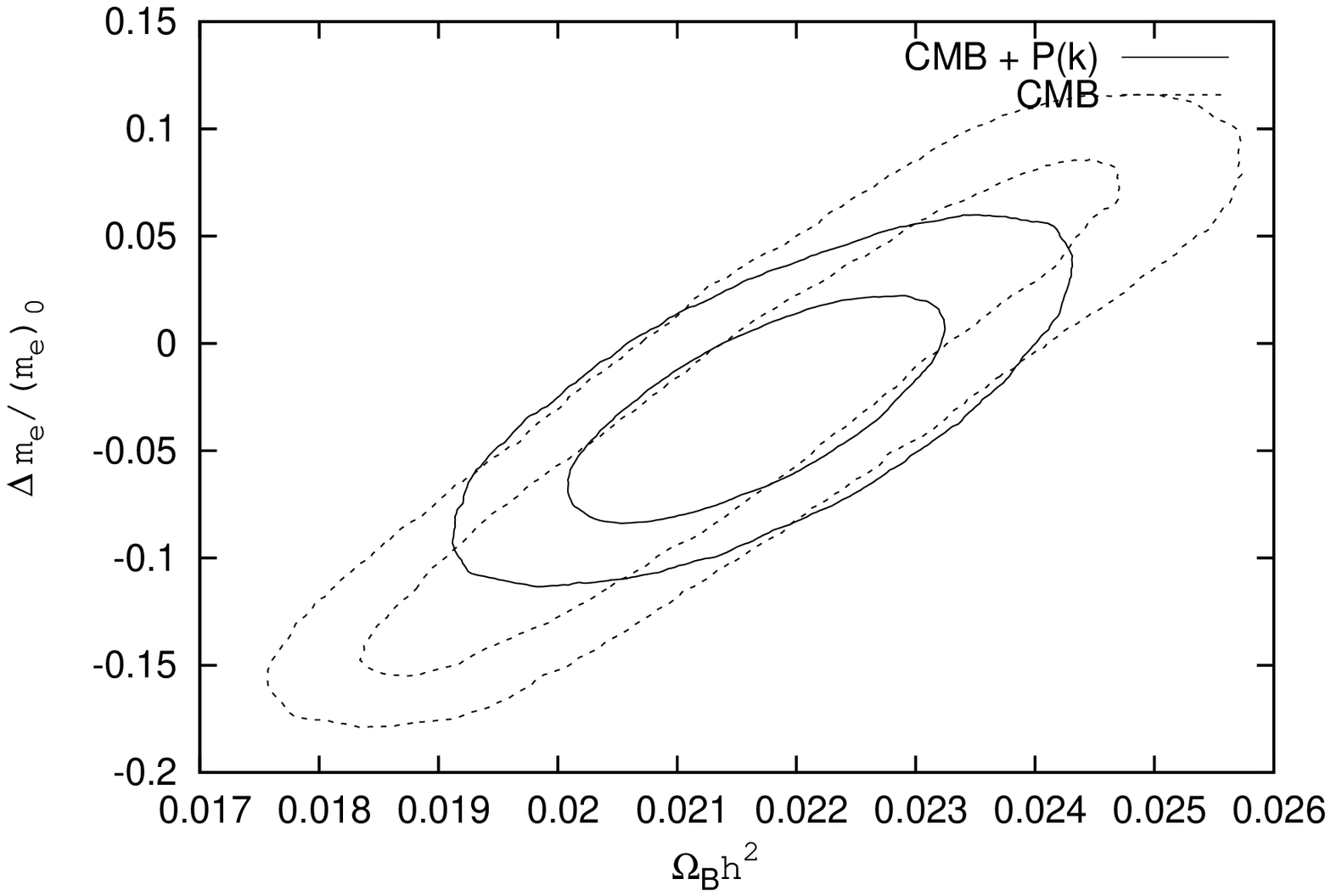}
\end{center}
\caption{1$\sigma$, and 2$\sigma$ confidence levels contours obtained with CMB
  data with and without data of the 2dFGRS power spectrum.}
\label{individuales}
\end{figure}

We have also performed the analysis considering only CMB data. The
strong degeneracy between $m_e$ and $H_0$ made the chains cover all
the wide $H_0$ prior, making it impossible to find reliable mean
values and errors. Hence, we added a gaussian prior to $H_0$, which
was obtained from the Hubble Space Telescope Key Project
\citep{hst01}, and chose the values of the mean and errors as those
inferred from the closest objects in that paper, so we could neglect
any possible difference between the value of $m_e$ at that redshift
and the present value.  In this way, we post-processed the chains and
found limits that are consistent with those of the first analysis,
revealing the robustness of these bounds. However, the most stringent
constraints were obtained in the first analysis (see figure
\ref{individuales}).

Finally, we comment that {\it Planck} will be the first mission
to map the entire CMB sky with mJy sensitivity and resolution better
than 10' \citep{Planck06}. Such resolution
will allow to see into the damping tail of the anisotropy spectrum,
around the third and fourth peaks, with a precision almost only
limited by cosmic variance \citep{White06}. This will enable a
very precise estimation of the baryon-to-photon ratio from the
relative height of the peaks in the spectrum.

\section{Bounds from Quasar Absorption Systems}
\label{quasars}
Quasar absorption systems present ideal laboratories to search for any
temporal variation of the fundamental constants over cosmological time
scales. In particular, a method for constraining the variation in $\mu
=\frac{m_p}{m_e}$ was developed by \citet{VL93}. It is based on the
fact that wavelengths of electron-vibro-rotational lines depend on the
reduced mass of the molecules, with different dependence for different
transitions. In such way, it is possible to distinguish the
cosmological redshift of a line from the shift caused by a variation
in $\mu$. The rest-frame laboratory wavelength, $\lambda_i^0$, can be
related to those in the quasar absorption system, $\lambda_i$, as
$\frac{\lambda_i}{\lambda_i^0} = (1 + z_{abs}) (1 + K_i \frac{\Delta
\mu}{\mu})$, where $z_{abs}$ is the absorption redshift and $K_i$ is
the coefficient which determines the sensitivity of the wavelength
$\lambda_i$. Using observations from ${\rm H}_2$ absorption systems at
high redshift and laboratory measurements, several authors obtained
constraints on $\mu$ \citep{P98,LE02,Ivanchik03,Ivanchik05}. The most
up-to-date available measurements for each redshift are listed in
table \ref{molecular} and will be considered to test Barrow-Magueijo
model.
\begin{table}[h!]
\begin{center}
\caption{The table shows the absorption redshift, the value of
  $\frac{\Delta m_e}{\left(m_e\right)_0}$  with its 
corresponding error (in units of $10^{-5}$), 
and the reference obtained comparing molecular and laboratory wavelengths.}
\label{molecular}
\begin{tabular}{|c|c|c|}
\hline
 Redshift  & $\frac{\Delta m_e}{\left(m_e\right)_0}\pm
 \sigma$  &  Reference \\ 
\hline
2.8 & $-6.25 \pm 13.70 $ &  \citet{P98}\\\hline
3.02 & $-1.40 \pm 0.83 $  & \citet{Ivanchik05}\\\hline
2.6 & $ -2.11 \pm 1.39$ &  \citet{Ivanchik05} \\\hline
\end{tabular}
\end{center}
\end{table}

Another method for constraining variation in fundamental constants is
based on the comparison between the hyperfine 21 cm absorption
transition of neutral hydrogen $(\nu _a)$ and an optical resonance
transition $(\nu _b)$. The ratio $\frac{\nu_a}{\nu_b}$ is proportional
to $x=\alpha ^2g_p\frac{m_e}{m_p}$ where $g_p$ is the proton $g$
factor \citep{Tzana07}. Thus, a change of this quantity will result in
a difference in the redshift measured from 21 cm and optical
absorption lines as $\frac{\Delta
x}x=\frac{z_{opt}-z_{21}}{\left(1+z_{21}\right) }$.  Since we are
working in the context of the \citet{BM05} model, the only fundamental
constant which is allowed to vary is $m_e$.  Table \ref{opticoradio}
shows the bounds obtained by \citet{Tzana07} combining the
measurements of optical and radio redshift. This method has the
inconvenience that it is difficult to determine if both radio and
optical lines were originated at the same absorption system. Thus, a
difference in the velocity of the absorption clouds could hide a
variation in $x$.
\begin{table}[h!]
\begin{center}
\caption{The table shows the absorption redshift, the value of
  $\frac{\Delta m_e}{\left(m_e\right)_0}$  with its 
corresponding error (in units of $10^{-5}$), obtained comparing radio
  and molecular redshifts \citep{Tzana07}.} 
\label{opticoradio}
\begin{tabular}{|c|c|c|c|c|}
\cline{1-2}\cline{4-5}
 Redshift  & $\frac{\Delta m_e}{\left(m_e\right)_0}\pm \sigma$
 &&Redshift  & $\frac{\Delta m_e}{\left(m_e\right)_0}\pm \sigma$ \\
\cline{1-2}\cline{4-5}
 0.24 &   $1.21 \pm   2.10$  &&  1.78 & $ -2.59 \pm  0.90$ \\ \cline{1-2}\cline{4-5}
  0.31 & $-0.61 \pm  4.27$  &&  1.94 & $ 3.30  \pm   0.44$ \\ \cline{1-2}\cline{4-5}
  0.40 & $ 3.22 \pm  3.15$ &&  2.04 & $ 5.20 \pm  2.76$    \\ \cline{1-2}\cline{4-5}
  0.52 & $-2.95 \pm  1.05 $ && 2.35 & $ -2.54 \pm  1.82$ \\ \cline{1-2}\cline{4-5} 
  0.52 & $  0.26 \pm 3.67 $ \\\cline{1-2}
 \end{tabular}
\end{center}
\end{table}

\section{Bounds from Geophysical Data}
\label{geo}
The half-life of long-lived $\beta $ decayers has been determined
either in laboratory measurements or by comparison with the age of
meteorites, as found from $\alpha $ decay radioactivity analysis.  
The most stringent bound on the variation in the half life, $\lambda$,
proceeds from the comparison of $^{187}\rm{Re}$ decay in the Solar 
System formation and the present \citep{Olive04b}: 
$\frac{\Delta \lambda}{\lambda} = (-0.016 \pm 0.016) $.
\citet{SV90} derived a relation between the
shift in the half-life of long lived $\beta $ decayers and a possible
variation between the values of the fundamental constants values now. 
In this paper, we only consider $m_e$ variation and
therefore $ \frac{\Delta
  \lambda}{\lambda} = a \frac{\Delta m_e}{\left(m_e\right)_0}$, where $a=-600$ for $^{187}\rm{Re}$.

\section{Bounds from Laboratory}
\label{clocks}
Comparison between frequencies of different atomic transitions over
time are useful tools to put stringent bounds on the variation in
fundamental constants at present. In particular, the hyperfine
frequency of cesium can be approximated by $\nu_{{\rm Cs}} \simeq g_{{\rm Cs}}
\frac{m_e}{m_p} \alpha^2 R_y F_{{\rm Cs}} (\alpha)$  (where $g_{{\rm Cs}}$ is the
nuclear $g$ factor, $R_y$ is the Rydberg constant expressed as a
frequency and $F_{{\rm Cs}}(\alpha)$ is a dimensionless function of
$\alpha$ and  does not depend on $m_e$ at least at first order), while
optical transition frequencies can be expressed as 
$\nu_{opt} \simeq R_y F(\alpha)$. 
Several authors \citep{Bize03,Fischer04,Peik04} have measured different optical
transitions and compared them with the frequency of the ground state
hyperfine splitting in neutral $^{133}{\rm Cs}$.  These measurements can be
used to constrain the variation in $\frac{\dot m_e}
{\left(m_e\right)_0}$. Constraints from different experiments are
listed in table \ref{table-clocks}.
\begin{table}[h!]
\begin{center}
\caption{The table shows the compared clocks, the value of
  $\frac{\dot m_e}{\left(m_e\right)_0}$ with its corresponding error
  (in units of $10^{-15}$), the time interval for which the variation was measured and the reference.}
\label{table-clocks}
\begin{tabular}{|c|c|c|c|}
\hline
 Frequencies  & $\frac{\dot m_e}{\left(m_e\right)_0} \pm \sigma
   \left[10^{-15} \rm{yr}^{-1}\right]$ &
 $\Delta t [\rm{yr}]$&  Reference \\ \hline
Hg$^{+}$ and Cs & $ 0.2 \pm 7.0 $ & 5  &  \citet{Fischer04}\\\hline
Yb$^+$ and Cs & $1.2 \pm 4.4 $ & 2.8 & \citet{Peik04}\\\hline
Hg$^+$ and Cs & $ 0 \pm 7 $ & 2 & \citet{Bize03} \\\hline
\end{tabular}
\end{center}
\end{table}

\section{The Model}
\label{modelo}
We now analyze the Barrow-Magueijo model for the variation in
$m_e$. We solve the equation of the scalar field $(\phi)$ that drives 
the variation in $m_e$ in this model. We consider that the variations
in $\phi$ are small and they do not produce significants contributions to
the Friedmann equation. As we did in a previous work
\citep{Mosquera07}, we build a piecewise approximate solution by
joining solutions obtained by keeping only some terms of the Friedmann
equation, relevant in the following domination regimes: a) radiation
and matter, and b) matter and cosmological constant. In such way,
solution a) can be applied to nucleosynthesis and recombination of
primordial hydrogen whereas solution b) is appropriate for quasar
absorption systems, geophysical data and atomic clocks.  

Defining the variable $\vartheta$ as $\rm{d}\vartheta=
d\tau/$$a$, where $\tau=H_0 t$, and $t$ is the cosmic time, the
expression for the scale factor in the radiation and matter regime is:
\begin{eqnarray}
\label{adeambos}
a_{RM}(\vartheta) &=&\frac{ \Omega_m}{4} \vartheta^2  + \sqrt{\Omega_r} \vartheta
\end{eqnarray}
and the relationship between $\tau$ and $\vartheta$ is: 
\begin{eqnarray}
\label{tdeambos} 
\tau(\vartheta)&=&\frac{\Omega_m}{12} \vartheta^3
 +\frac{ \sqrt{\Omega_r}}{2}\vartheta^2
\end{eqnarray}

The solution for the scale factor in the matter and cosmological
constant regime can be written as:
\begin{eqnarray}
\label{amyc} a_{MC}(\tau) &=& \left(
\frac{\Omega_m}{\Omega_\Lambda}\right)^{1/3} \left[ {\rm sinh}
\left( \frac{3}{2} \sqrt{\Omega_\Lambda} \left(\tau-\tau_0 \right)+
\sinh^{-1} \sqrt{\frac{\Omega_\Lambda}{\Omega_m}}
 \right)\right]^{2/3}
\end{eqnarray}
where $\tau_0= H_0t_0$, and $t_0$ is the age of the universe. To
obtain the last solution we have considered that the scale factor
must be a continuous and smooth function of time.

In the Barrow-Magueijo model, $m_e$ is controlled by a
dilaton field $\phi$ defined by $m_e=(m_e)_0\exp (\phi)$, and variations
in $m_e$ occur no sooner the universe cools down below $m_e$
threshold. The minimal dynamics for $\phi$ is set by the 
kinetic Lagrangian
\begin{equation}
\mathcal{L}_{\phi }={\frac{\omega }{2}}\partial _{\mu }\phi \partial ^{\mu
}\phi
\end{equation}
where $\omega$ is a coupling constant.  From this Lagrangian, the
equation of motion of the scalar field can be derived as:
\begin{equation}
(\dot{\phi}a^{3}\dot{)}=-M\exp [\phi ]  \label{eqM}
\end{equation}
 with $M\simeq \rho_{e0}a_0^3 c^4 /\omega$. This is a second order
equation for $\phi$, with the boundary condition $\phi_0=0$.  If the
mass variations are small, $e^\phi \simeq 1$ can be set to obtain an
analytical expression for $\phi$.

For convenience, we define $\beta= \frac{1}{4} \OmegaM
\OmegaR^{-1/2}$, $\xi= \OmegaM \OmegaR^{-3/4}\OmegaL^{-1/4}$,
$\gamma=\OmegaL^{1/2}\OmegaM^{-1/2}$ and
\begin{eqnarray}
f(\xi) &=& \frac{2+(\xi-2)\sqrt{1+\xi}}{\xi^2} \nonumber \\
C &=& \sinh^{-1}{\xi^{-1/2} - f(\xi)} 
\end{eqnarray}
Provided $\phi= \ln (m_e/(m_e)_0)\simeq \Delta m_e/(m_e)_0$, the
expressions for the variation in $m_e$ in the two regimes
are:
\begin{itemize}
\item for $\tau<\tau_1$ (where $\tau_1$ is defined by
  $a_{RM}(\tau_1)=a_{MC}(\tau_1)=\left(\frac{\Omega_r}{\Omega_\Lambda}\right)^{1/4}$, see
  \citet{Mosquera07}):
\begin{eqnarray}
\frac{\Delta m_e}{(m_e)_0}(\vartheta)&=& \frac{2}{3} \frac{M}{H_0^2}
  \frac{1}{\OmegaM}\left[ -2 \ln \left(\frac{2(\beta \vartheta + 1)}{1+\sqrt{1+\xi}} \right)
+\frac{1}{\beta\vartheta +1} -\frac{2}{1+\sqrt{1+\xi}} +
  \frac{2}{3}f(\xi)\sqrt{1+\xi}\right. \nonumber \\ 
&& \hskip 2cm\left.   +\frac{1}{4}\ln \left(\frac{\OmegaL}{\OmegaR}\right)
  -\frac{2}{3} \left( \sinh^{-1}\gamma - C \right)
  \frac{\sqrt{1+\gamma^2}}{\gamma} \right]  \nonumber \\
&&+ \frac{A}{H_0}
  \frac{\OmegaM}{\OmegaR^{3/2}} \left[ \frac{1}{2}\ln
  \left(\frac{\beta \vartheta +1}{\beta \vartheta}\right) 
+ \frac{1}{2}\ln\left(\frac{\sqrt{1+\xi} -1}{\sqrt{1+\xi} +1}\right)
  -\frac{1}{4\beta \vartheta}  -\frac{1}{4(\beta \vartheta +1)}
  \right. \nonumber \\
&&\hskip 2cm \left. +  \frac{\left( \xi - \frac{2}{3}\right)\sqrt{1+\xi} +
  \frac{2}{3}\frac{\sqrt{1+\gamma^2}}{\gamma}}{\xi^2} \right]   
\end{eqnarray}
The relationship between $\tau$ and $\vartheta$ is given by Eq.(\ref{tdeambos}).
\item for $\tau>\tau_1$
\begin{eqnarray}
\frac{\Delta m_e}{(m_e)_0}(\tau) &= &\phi_0 +
  \frac{M}{H_0^2}\, \frac{2}{3\OmegaM}\left[ \sqrt{\OmegaL} \tau \coth
    \left( C +  \frac{3}{2}\sqrt{\OmegaL}\tau  \right)  - \frac{2}{3} \ln \left[\sinh \left( C +
  \frac{3}{2}\sqrt{\OmegaL}\tau \right) \right] \right.
    \nonumber\\ 
&& \hskip 2.5cm\left. +  \frac{2}{3}  \left(
  \ln\gamma - \frac{\sqrt{1 + \gamma^2}}{\gamma} \left[ C + \ln
  \left(\gamma + \sqrt{1+\gamma^2}\right) \right]\right)   \right] +
  \nonumber\\ 
&& +
  \frac{A}{H_0}\,\frac{2\sqrt{\OmegaL}}{3\OmegaM}\left[-\coth\left(C +
  \frac{3}{2}\sqrt{\OmegaL}\tau \right) +   \frac{\sqrt{1 +
  \gamma^2}}{\gamma} \right] 
\end{eqnarray}
\end{itemize}
where $A$ is an integration constant.

\section{Results}
\label{resultados}
The model described in section \ref{modelo} predicts the variation in
$m_e$ as a function of time, and has two independent
dimensionless parameters $M/H_0^2$ and $A/H_0$. We do not fix $A/H_0$
to zero as previous works did \citep{BM05}. To constrain these
parameters, we use the data described in the previous sections.  We
perform a $\chi^2$ test to obtain the best fit parameters of the
model. In order to obtain the parameters consistently with our
assumption that the energy density of the field $\phi$
$\left(\epsilon_\phi = \frac{1}{c^2}\frac{\omega}{2}\dot
\phi^2\right)$ can be neglected in the Friedmann equation, we add to
the $\chi^2$ expression, a term that controls that the contribution of
$\phi$ to the Friedmann equation will be less important that the
radiation term, right after $m_e$ threshold is crossed. 
The result of the statistical analysis shows that there is no good
fit for the whole data set. We repeat the 
analysis excluding one group of data at each time. We found that
reasonable fits can be obtained excluding the quasar at $z=1.94$ of table
\ref{opticoradio} and that the data from nucleosynthesis is crucial to
determine the value of $A/H_0$. Besides, the group of data from table
\ref{molecular} is important to determine the value of $M/H_0^2$.
The results are shown in table \ref{table-results}. 
\begin{table}[h!]
\caption{The table shows the best fit parameters of the model 
(excluding entry 7 of table \ref{opticoradio}). The value for the
 $M/H_0^2$ parameter is in units of $10^{-6}$, and the value for
  the $A/H_0$ parameter is in units of $10^{-13}$.}  
\begin{center}
\label{table-results}
\begin{tabular}{|c|c|c|c|}
\hline
 $M/H_0^2 $ & $A/H_0$ & $G\omega /c^4$  &$\frac{\chi^2_{min}}{N-2}$  \\ \hline 
  $-7.30_{-2.02}^{+2.10}$   &
 $3.60_{-1.50}^{+1.44}$  & $-0.336_{-0.093}^{+0.097}$ & 1.14  \\ \hline 
\end{tabular}
\end{center}
\end{table}

Since the Barrow-Magueijo model is written in terms of the coupling constant
$\omega$, we derive its best value from the previous constraints.
Since $M\simeq \rho_{e0}a_0^3 c^4/\omega$ 
we obtain the following relationship:
\begin{equation}
\frac{G \omega}{c^4}= \frac{3}{8\pi} \ \left(1 - \frac{f_{{\rm He}}}{2}\right)\ \Omega_b\ \frac{m_e}{m_p}
\ \left(\frac{M}{H_0^2}\right)^{-1}
\end{equation}
where $f_{{\rm He}}$ is the fraction of the total number of baryons in the
form of ${\rm He}$, and can be written as a function of the total observed
mass abundance of ${\rm He}$ ($M_{{\rm He}}/M_{\rm H}$). According to the values of $M/H_0^2$ from table
\ref{table-results} and using $f_{{\rm He}}=0.19$ (taking
$M_{{\rm He}}/M_{\rm H}=0.24$), we obtain the bounds on the dimensionless quantity
$\frac{G \omega}{c^4}$ presented in table \ref{table-results}.

\section{Summary and Conclusions}
\label{resumen}
In this paper, we have put limits on the time variation in the
electron mass at primordial nucleosynthesis time using observational
primordial abundances of $\De$, $\He$ and $\Li$, and we have
analyzed in detail the consequences of considering different groups of
data. We have also considered the variations in $<v>$ during BBN and 
analyzed the differences with the variations in $m_e$ during the same epoch.
Additionally, we have used the three year data from the Cosmic
Microwave Background and the final 2dFGRS power spectrum to obtain
bounds on the variation in $m_e$ at recombination, and an
estimation of the cosmological parameters. Together with other bounds
on the variation in the late universe, that come from quasar
absorption systems, half-life of long-lived $\beta$ decayers, and
atomic clocks, we put constraints on the Barrow-Magueijo model for the
variation in $m_e$. We have improved the solutions by
taking into account the detailed evolution of the scale factor and
the complete solution for the scalar field that drives the variation
in $m_e$.

In the original paper \citep{BM05} some approximations in the evolution of the
scale factor are assumed with the consequent simplification in the
solution for the scalar field. Another improvement of our derivation
is that we have not neglected the first integration constant, which is
the most contributing part in the early universe. In fact, integrating
Eq.(\ref{eqM}) once, we can write:
\begin{equation}
\dot{\phi}a^{3}=-M\left( t - \frac{A}{M}\right) 
=-M\left( t + 8.47\times 10^{2} {\rm yr} \right) \label{eqM2}
\end{equation}
where we have used the best fit values for the parameters $M/H_0^2$
and $A/H_0$, and $h=0.73$. Note that the second term in the right hand
side of Eq.(\ref{eqM2}) is dominant in the early universe, in
particular, during nucleosynthesis.

\citet{BM05} presented a bound of $ G|\omega| > 0.2$ (with $c=1$). They
obtained such constraint using bounds from quasars at $z\sim 1$, whereas
we use all the available bounds on the variation in $m_e$ at different
cosmological times. In appendix \ref{apendice} we briefly discuss the
difference in both analysis. From data supporting the
weak equivalence principle, they obtain $G|\omega| > 10^3$ while we obtain
$-0.615 < G\omega/c^4 < -0.045 $ (3$\sigma$ interval) using data from different
cosmological time scales. More research both on time variation
data and on the bound from WEP is needed to understand this
discrepancy.

Finally, we remark that, 
at 2$\sigma$, the value of $\omega$ is negative. This should not be
surprising. Indeed, negative kinetic terms in the Lagrangian have
already been considered in k-essence models with a phantom energy
component \citep{caldwell02}.

\section*{{\bf Acknowledgements}}
Support for this work was provided by Project G11/G071, UNLP and PIP
5284 CONICET.  The authors would like to thank Andrea Barral, Federico
Bareilles, Alberto Camyayi and Juan Veliz for technical and
computational support. The authors would also like to thank Ariel
Sanchez for support with CosmoMC. MEM wants to thank Sergio Iguri for
the helpful discussions. CGS gives special thanks to Licia Verde and
Nelson Padilla for useful discussion. SJL wants to thank Luis Chimento
for useful discussions.

\appendix

\section{Physics at BBN}
\label{correccion}
In this appendix we discuss the dependences of the physical quantities
involved in the calculation of the abundances of the light elements
with $m_e$ and variation in the Higgs vacuum  
expectation value. We also discuss how this quantities are changed within the Kawano Code.

\subsection{Variation in the electron mass}
A change in the value of $m_e$ at the time of primordial
nucleosynthesis with respect to its present value affects derived
physical quantities such as  the sum of the electron and positron
energy densities, the sum of the 
electron and positron pressures and the difference of the electron and positron 
number densities. In Kawano's code, these quantities are calculated as follows:
\begin{eqnarray}
\label{rhoe}
\rho_{e^-}+ \rho_{e^+}&=& \frac{2}{\pi^2} \frac{\left(m_e
  c^2\right)^4}{\left(\hbar c \right)^3} \sum_n (-1)^{n+1} {\rm ch}
\left(n\phi_e\right) M(nz)\\
\label{pe}
\frac{p_{e^-}+ p_{e^+}}{c^2}&=& \frac{2}{\pi^2} \frac{\left(m_e
  c^2\right)^4}{\left(\hbar c \right)^3} \sum_n \frac{(-1)^{n+1}}{nz} {\rm ch}
\left(n\phi_e\right) N(nz)\\
\label{ne}
 \frac{\pi^2}{2}\left[\frac{\hbar c^3}{m_e c^2}\right]^3
   z^3\left(n_{e^-}-n_{e^+}\right)&=& z^3 \sum_n (-1)^{n+1}{\rm
     sh}\left(n\phi_e\right) L(nz)
\end{eqnarray}
where $z=\frac{m_e c 2}{k T_\gamma}$, $\phi_e$ is the electron
chemical potential and $L(z)$,  $M(z)$, and $N(z)$ are related to the modified Bessel function
\citep{Kawano88,Kawano92}. In order to include the variation in $m_e$
we replace, in all the equations, $m_e$ by 
$\left(m_e\right)_0 \left(1+\frac{\Delta m_e}{\left(m_e\right)_0} \right)$.
The change in these quantities, due to a change in $m_e$, 
affects their derivatives and the expansion rate through the
Friedmann equation.
The $n \leftrightarrow p$ reaction rates and the other weak decay
rates are changed if $m_e $ varies with time. The total $n\rightarrow p$
reaction rate is calculated by: 
\begin{eqnarray} 
\label{lambdanp}
\!\!
\lambda=
K \int_{m_e}^{\infty}  \frac{ {\rm d}E_e \, \,  E_e
  p_e \left(E_e + \Delta
  m_{np}\right)^2}{\left(1+e^{E_e/T_\gamma}\right)\left(1+e^{-\left(E_e + \Delta m_{np}\right)/T_\nu-\xi_l}\right)}
+ K \int_{m_e}^{\infty} \frac{{\rm d}E_e \, \, E_e p_e \left(E_e - \Delta m_{np}\right)^2}{\left(1+e^{-E_e/T_\gamma}\right)\left(1+e^{\left(E_e - \Delta m_{np}\right)/T_\nu -\xi_l}\right)} \,\,
\end{eqnarray}
where $E_e$ and $p_e$ are the electron energy and momentum respectively, $\Delta m_{np}$ is the neutron-proton mass difererences, $K$ is a normalization
constant proportional to $G_F^2$, and $E_e = \left(p_e^2 +
m_e^2\right)^{1/2}$. 

It is worth to mention that the most important changes in the
primordial abundances (due to a change in $m_e$) arrives from the
change in the weak rates rather than the change in the expansion rate
\citep{YS03}. 

\subsection{Variation in the Higgs vacuum expectation value}
If the the value of $<v>$ during BBN is different than its present value, 
the electron mass, the Fermi constant, the neutron-proton mass difference and the
deuterium binding energy take different values than the current ones. The electron mass
is proportional to the Higgs vacuum expectation value.
The Fermi constant is proportional to $<v>^{-2}$ \citep{dixit88}. This dependence affect the 
$n\leftrightarrow p$ reaction rates since $K\sim G_F^2$.
The neutron-proton mass difference changes by \citep{Epele91b}
\begin{eqnarray}
\frac{\delta \Delta m_{np}}{\Delta m_{np}} &=&1.587 \frac{\Delta <v>}{<v>_0}
\end{eqnarray}
affecting $n\leftrightarrow p$ reaction rates (see
Eq.(\ref{lambdanp})) and the initial neutrons and protons abundances: 
\begin{eqnarray}
Y_n= \frac{1}{1+e^{\Delta m_{np}/T_9 +\xi}}\hskip 2cm
Y_p=  \frac{1}{1+e^{-\Delta m_{np}/T_9 -\xi}}
\end{eqnarray}
where $T_9$ is the temperature in units of $10^9 \,{\rm K}$.
In order to include these effects we replace $\Delta m_{np}$ by $\Delta
m_{np}\left(1+\frac{\delta \Delta m_{np}}{\Delta m_{np}}\right)$. 
The deuterium binding energy must also be corrected by 
$\frac{\Delta \epsilon_D}{\epsilon_D} =k \frac{\Delta <v>}{<v>_0}$
 where $k$ is a model dependent constant. In this work we assume,
following \cite{Chamoun07}, $k=-0.045$. 
This correction affects the initial value of the deuterium abundance.
Once again we replace $\epsilon_D$ by
$\epsilon_D \left(1+\frac{\Delta \epsilon_D}{\epsilon_D}\right)$ in
the code.

\section{Physics at recombination epoch}
\label{apendice_recombinacion}
During recombination epoch, the ionization fraction, $x_e$, is determined by the balance
between photoionization
and recombination. The recombination equation is
\begin{equation}
-\frac{d}{dt}\left(\frac{n_e}{n}\right) = C \left( \frac{\alpha_c
 n_e^2}{n} -\beta_c \frac{n_{1s}}{n} e^{-(B_1 - B_2)/kT} \right)
\end{equation}
where 
\begin{equation}
C= \frac{\left(1 + K \Lambda_{2s,1s} n_{1s}\right)}{\left(1
 + K (\beta_c + \Lambda_{2s,1s}) n_{1s}\right)}
\end{equation}
is the Peebles factor, which inhibits the recombination rate due to
the presence of Lyman-$\alpha$ photons. The redshift of these photons
is $K = \frac{\lambda_\alpha^3 \ a}{8 \pi \dot a}$, with
$\lambda_\alpha = \frac{8 \pi \hslash c}{3 B_1}$, and
$\Lambda_{2s,1s}$ is the rate of decay of the $2s$ excited state to
the ground state via 2-photon emission, and scales as $m_e$.
Recombination directly to the ground state is strongly inhibited, so
the \emph{case B} recombination takes place. The \emph{case B}
recombination coefficient $\alpha_c$ is proportional to 
$ m_e^{-3/2}$. The photoionization coefficient depends on $\alpha_c$,
but it also has an additional dependence on $m_e$,
\begin{equation}
\beta_c = \alpha_c \left(\frac{2 \pi {m_e} k
  T_m}{h^2}\right)^{3/2} e^{-B_2 / kT_m}
\nonumber 
\end{equation}
The most important effects of a change in $m_e$ during
recombination would be due to its influence upon Thomson scattering
cross section $\sigma _T = \frac{8\pi \ \hbar ^2c^2}{3\
  m_e^2}\alpha^2$, and the binding energy of hydrogen $B_1={1\over2}\ \alpha^2  m_e c^2$. 

\section{Different limits on $G\omega$}
\label{apendice}
In this appendix we compare the limits obtained by \citet{BM05} on $G
\omega$ with our bounds. 
We stress that we have performed a $\chi^2$ using
all available observational and experimental data, while \citet{BM05}
consider $|\frac{\Delta \mu}{\mu}|<10^{-5}$ for data at redshift of
order 1. Moreover, most exact individual bounds from quasar
absorption systems are not consistent with null variation at least at
$1\sigma$. 

Let us consider for example the last entry of table
\ref{molecular}: $-3.5 \times 10^{-5}< \frac{\Delta \mu}{\mu} < -0.72
\times 10^{-5} $. Using the same approximation as \citet{BM05}, we
find that $-0.28 < G \omega < -0.05 $ for this measurement which is of
the same order of magnitude as obtained considering all data.

\bibliography{bibliografia4} 

\begin{thebibliography}{}

\bibitem[\protect\astroncite{{Asplund et al.}}{2006}]{Asplund05}
{Asplund et al.}: 2006,
\newblock {\em \apj} {\bf 644}, 229

\bibitem[\protect\astroncite{{Barger et al.}}{2003}]{Barger03}
{Barger et al.}: 2003,
\newblock {\em Physics Letters B} {\bf 569}, 123

\bibitem[\protect\astroncite{{Barr} and {Mohapatra}}{1988}]{Barr88}
{Barr}, S.~M. and {Mohapatra}, P.~K.: 1988,
\newblock {\em Phys.Rev.D} {\bf 38}, 3011

\bibitem[\protect\astroncite{{Barrow} and {Magueijo}}{2005}]{BM05}
{Barrow}, J.~D. and {Magueijo}, J.: 2005,
\newblock {\em \prd} {\bf 72(4)}, 043521

\bibitem[\protect\astroncite{{Barrow} et~al.}{2002}]{BSM02}
{Barrow}, J.~D., {Sandvik}, H.~B., and {Magueijo}, J.: 2002,
\newblock {\em Phys.Rev.D} {\bf 65}, 063504

\bibitem[\protect\astroncite{Bekenstein}{1982}]{Bekenstein82}
Bekenstein, J.~D.: 1982,
\newblock {\em Phys.Rev.D} {\bf 25}, 1527

\bibitem[\protect\astroncite{{Bekenstein}}{2002}]{Bekenstein2002}
{Bekenstein}, J.~D.: 2002,
\newblock {\em Phys.Rev.D} {\bf 66}, 123514

\bibitem[\protect\astroncite{{Bergstr{\" o}m} et~al.}{1999}]{Iguri99}
{Bergstr{\" o}m}, L., {Iguri}, S., and {Rubinstein}, H.: 1999,
\newblock {\em Phys.Rev.D} {\bf 60}, 45005

\bibitem[\protect\astroncite{{Bize et al.}}{2003}]{Bize03}
{Bize et al.}: 2003,
\newblock {\em Physical Review Letters} {\bf 90(15)}, 150802

\bibitem[\protect\astroncite{{Boesgaard} et~al.}{2005}]{BNS05}
{Boesgaard}, A.~M., {Novicki}, M.~C., and {Stephens}, A.: 2005,
\newblock in V. {Hill}, P. {Francois}, and F. {Primas} (eds.), {\em Proceedings
  of IAU Symposium No. 228: "From Lithium to Uranium: Elemental Tracers of
  Early Cosmic Evolution"}, p.~29, Cambridge University Press

\bibitem[\protect\astroncite{{Bonifacio} and {Molaro}}{1997}]{bonifacio2}
{Bonifacio}, P. and {Molaro}, P.: 1997,
\newblock {\em Mon.Not.R.Astron.Soc.} {\bf 285}, 847

\bibitem[\protect\astroncite{{Bonifacio} et~al.}{1997}]{bonifacio1}
{Bonifacio}, P., {Molaro}, P., and {Pasquini}, L.: 1997,
\newblock {\em Mon.Not.R.Astron.Soc.} {\bf 292}, L1

\bibitem[\protect\astroncite{{Bonifacio et al.}}{2002}]{bonifacio3}
{Bonifacio et al.}: 2002,
\newblock {\em Astronomy and Astrophysics} {\bf 390}, 91

\bibitem[\protect\astroncite{{Bonifacio et al.}}{2007}]{bonifacio07}
{Bonifacio et al.}: 2007,
\newblock {\em \aap} {\bf 462}, 851

\bibitem[\protect\astroncite{{Brax et al.}}{2003}]{branes03b}
{Brax et al.}: 2003,
\newblock {\em Astrophysics and Space Science} {\bf 283}, 627

\bibitem[\protect\astroncite{{Burles} and {Tytler}}{1998a}]{burles1}
{Burles}, S. and {Tytler}, D.: 1998a,
\newblock {\em Astrophys.J.} {\bf 499}, 699

\bibitem[\protect\astroncite{{Burles} and {Tytler}}{1998b}]{burles2}
{Burles}, S. and {Tytler}, D.: 1998b,
\newblock {\em Astrophys.J.} {\bf 507}, 732

\bibitem[\protect\astroncite{{Caldwell}}{2002}]{caldwell02}
{Caldwell}, R.~R.: 2002,
\newblock {\em Physics Letters B} {\bf 545}, 23

\bibitem[\protect\astroncite{{Chamoun et al.}}{2007}]{Chamoun07}
{Chamoun et al.}: 2007,
\newblock {\em Journal of Physics G Nuclear Physics} {\bf 34}, 163

\bibitem[\protect\astroncite{{Christiansen et al.}}{1991}]{Epele91b}
{Christiansen et al.}: 1991,
\newblock {\em Physics Letters B} {\bf 267}, 164

\bibitem[\protect\astroncite{Cole et~al.}{2005}]{2dF05}
Cole, S. et~al.: 2005,
\newblock {\em Mon. Not. Roy. Astron. Soc.} {\bf 362}, 505

\bibitem[\protect\astroncite{{Crighton et al.}}{2004}]{Crighton04}
{Crighton et al.}: 2004,
\newblock {\em Mon.Not.R.Astron.Soc.} {\bf 355}, 1042

\bibitem[\protect\astroncite{{Damour} et~al.}{2002a}]{DPV2002a}
{Damour}, T., {Piazza}, F., and {Veneziano}, G.: 2002a,
\newblock {\em Physical Review Letters} {\bf 89}, 081601

\bibitem[\protect\astroncite{{Damour} et~al.}{2002b}]{DPV2002b}
{Damour}, T., {Piazza}, F., and {Veneziano}, G.: 2002b,
\newblock {\em Phys.Rev.D} {\bf 66}, 046007

\bibitem[\protect\astroncite{{Damour} and {Polyakov}}{1994}]{DP94}
{Damour}, T. and {Polyakov}, A.~M.: 1994,
\newblock {\em Nuclear Physics B} {\bf 95}, 10347

\bibitem[\protect\astroncite{{Dixit} and {Sher}}{1988}]{dixit88}
{Dixit}, V.~V. and {Sher}, M.: 1988,
\newblock {\em Phys.Rev.D} {\bf 37}, 1097

\bibitem[\protect\astroncite{{Fischer et al.}}{2004}]{Fischer04}
{Fischer et al.}: 2004,
\newblock {\em Physical Review Letters} {\bf 92(23)}, 230802

\bibitem[\protect\astroncite{{Freedman et al.}}{2001}]{hst01}
{Freedman et al.}: 2001,
\newblock {\em Astrophys.J.} {\bf 553}, 47

\bibitem[\protect\astroncite{{Gleiser} and {Taylor}}{1985}]{GT85}
{Gleiser}, M. and {Taylor}, J.~G.: 1985,
\newblock {\em Phys.Rev.D} {\bf 31}, 1904

\bibitem[\protect\astroncite{{Ivanchik et al.}}{2003}]{Ivanchik03}
{Ivanchik et al.}: 2003,
\newblock {\em Astrophysics and Space Science} {\bf 283}, 583

\bibitem[\protect\astroncite{{Ivanchik et al.}}{2005}]{Ivanchik05}
{Ivanchik et al.}: 2005,
\newblock {\em \aap} {\bf 440}, 45

\bibitem[\protect\astroncite{{Izotov} et~al.}{2007}]{izotov07}
{Izotov}, Y.~I., {Thuan}, T.~X., and {Stasi{\'n}ska}, G.: 2007,
\newblock {\em \apj} {\bf 662}, 15

\bibitem[\protect\astroncite{Jones et~al.}{2006}]{BOOM05_temp}
Jones, W.~C. et~al.: 2006,
\newblock {\em \apj} {\bf 647}, 823

\bibitem[\protect\astroncite{Kaluza}{1921}]{Kaluza}
Kaluza, T.: 1921,
\newblock {\em Sitzungber. Preuss. Akad. Wiss.K} {\bf 1}, 966

\bibitem[\protect\astroncite{{Kawano}}{1988}]{Kawano88}
{Kawano}, L.: 1988,
\newblock FERMILAB-PUB-88-034-A

\bibitem[\protect\astroncite{Kawano}{1992}]{Kawano92}
Kawano, L.: 1992,
\newblock FERMILAB-PUB-92-004-A

\bibitem[\protect\astroncite{{Kirkman et al.}}{2003}]{kirkman}
{Kirkman et al.}: 2003,
\newblock {\em Astrophys.J.Suppl.Ser.} {\bf 149}, 1

\bibitem[\protect\astroncite{Klein}{1926}]{Klein}
Klein, O.: 1926,
\newblock {\em Z. Phys.} {\bf 37}, 895

\bibitem[\protect\astroncite{{Kujat} and {Scherrer}}{2000}]{KS00}
{Kujat}, J. and {Scherrer}, R.~J.: 2000,
\newblock {\em \prd} {\bf 62(2)}, 023510

\bibitem[\protect\astroncite{Kuo et~al.}{2004}]{ACBAR02}
Kuo, C.-l. et~al.: 2004,
\newblock {\em Astrophys. J.} {\bf 600}, 32

\bibitem[\protect\astroncite{{Levshakov et al.}}{2002}]{LE02}
{Levshakov et al.}: 2002,
\newblock {\em Mon.Not.R.Astron.Soc.} {\bf 333}, 373

\bibitem[\protect\astroncite{{Lewis} and {Bridle}}{2002}]{LB02}
{Lewis}, A. and {Bridle}, S.: 2002,
\newblock {\em \prd} {\bf 66(10)}, 103511

\bibitem[\protect\astroncite{{Lewis} et~al.}{2000}]{LCL00}
{Lewis}, A., {Challinor}, A., and {Lasenby}, A.: 2000,
\newblock {\em \apj} {\bf 538}, 473

\bibitem[\protect\astroncite{{Maeda}}{1988}]{Maeda88}
{Maeda}, K.: 1988,
\newblock {\em Modern Physics. Letters A} {\bf 31}, 243

\bibitem[\protect\astroncite{{Mosquera et al.}}{2007}]{Mosquera07}
{Mosquera et al.}: 2007,
\newblock {\em ArXiv e-prints}

\bibitem[\protect\astroncite{{Olive et al.}}{2004}]{Olive04b}
{Olive et al.}: 2004,
\newblock {\em \prd} {\bf 69(2)}, 027701

\bibitem[\protect\astroncite{{Oliveira et al.}}{2006}]{oliveira06}
{Oliveira et al.}: 2006,
\newblock {\em \apj} {\bf 642}, 283

\bibitem[\protect\astroncite{{O'Meara et al.}}{2001}]{omeara}
{O'Meara et al.}: 2001,
\newblock {\em Astrophys.J.} {\bf 552}, 718

\bibitem[\protect\astroncite{{O'Meara et al.}}{2006}]{omeara06}
{O'Meara et al.}: 2006,
\newblock {\em \apjl} {\bf 649}, L61

\bibitem[\protect\astroncite{{Overduin} and {Wesson}}{1997}]{OW97}
{Overduin}, J.~M. and {Wesson}, P.~S.: 1997,
\newblock {\em Phys.Rep.} {\bf 283}, 303

\bibitem[\protect\astroncite{{Palma et al.}}{2003}]{branes03a}
{Palma et al.}: 2003,
\newblock {\em Phys.Rev.D} {\bf 68}, 123519

\bibitem[\protect\astroncite{{Peik et al.}}{2004}]{Peik04}
{Peik et al.}: 2004,
\newblock {\em Physical Review Letters} {\bf 93(17)}, 170801

\bibitem[\protect\astroncite{{Peimbert} et~al.}{2007}]{PL07}
{Peimbert}, M., {Luridiana}, V., and {Peimbert}, A.: 2007,
\newblock {\em \apj} {\bf 666}, 636

\bibitem[\protect\astroncite{{Pettini} and {Bowen}}{2001}]{pettini}
{Pettini}, M. and {Bowen}, D.~V.: 2001,
\newblock {\em Astrophys.J.} {\bf 560}, 41

\bibitem[\protect\astroncite{Piacentini et~al.}{2006}]{BOOM05_polar}
Piacentini, F. et~al.: 2006,
\newblock {\em \apj} {\bf 647}, 833

\bibitem[\protect\astroncite{{Potekhin et al.}}{1998}]{P98}
{Potekhin et al.}: 1998,
\newblock {\em Astrophys.J.} {\bf 505}, 523

\bibitem[\protect\astroncite{Raftery and Lewis}{1992}]{Raftery&Lewis}
Raftery, A.~E. and Lewis, S.~M.: 1992,
\newblock in J.~M. Bernado (ed.), {\em Bayesian Statistics}, p. 765, OUP

\bibitem[\protect\astroncite{Readhead et~al.}{2004}]{CBI04}
Readhead, A. C.~S. et~al.: 2004,
\newblock {\em Astrophys. J.} {\bf 609}, 498

\bibitem[\protect\astroncite{{Ryan et al.}}{2000}]{ryan}
{Ryan et al.}: 2000,
\newblock {\em Astrophys.J.} {\bf 530}, L57

\bibitem[\protect\astroncite{{Sanchez et al.}}{2006}]{Sanchez06}
{Sanchez et al.}: 2006,
\newblock {\em Mon. Not. Roy. Astron. Soc.} {\bf 366}, 189

\bibitem[\protect\astroncite{{Seager} et~al.}{1999}]{recfast}
{Seager}, S., {Sasselov}, D.~D., and {Scott}, D.: 1999,
\newblock {\em \apjl} {\bf 523}, L1

\bibitem[\protect\astroncite{{Sisterna} and {Vucetich}}{1990}]{SV90}
{Sisterna}, P. and {Vucetich}, H.: 1990,
\newblock {\em Phys.Rev.D} {\bf 41}, 1034

\bibitem[\protect\astroncite{{Spergel et al.}}{2003}]{wmapest}
{Spergel et al.}: 2003,
\newblock {\em Astrophys.J.Suppl.Ser.} {\bf 148}, 175

\bibitem[\protect\astroncite{{Spergel et al.}}{2007}]{wmap3}
{Spergel et al.}: 2007,
\newblock {\em \apjs} {\bf 170}, 377

\bibitem[\protect\astroncite{{Steigman}}{2005}]{Steigman05}
{Steigman}, G.: 2005,
\newblock {\em Physica Scripta Volume T} {\bf 121}, 142

\bibitem[\protect\astroncite{{Steigman}}{2006}]{Steigman06}
{Steigman}, G.: 2006,
\newblock {\em International Journal of Modern Physics E} {\bf 15}, 1

\bibitem[\protect\astroncite{{The Planck Collaboration}}{2006}]{Planck06}
{The Planck Collaboration}: 2006,
\newblock {\em ArXiv Astrophysics e-prints}

\bibitem[\protect\astroncite{{Tzanavaris et al.}}{2007}]{Tzana07}
{Tzanavaris et al.}: 2007,
\newblock {\em \mnras} {\bf 374}, 634

\bibitem[\protect\astroncite{{Varshalovich} and {Levshakov}}{1993}]{VL93}
{Varshalovich}, D.~A. and {Levshakov}, S.~A.: 1993,
\newblock {\em Soviet Journal of Experimental and Theoretical Physics Letters}
  {\bf 58}, 237

\bibitem[\protect\astroncite{{Weinberg}}{1983}]{Weinberg83}
{Weinberg}, S.: 1983,
\newblock {\em Physics Letters B} {\bf 125}, 265

\bibitem[\protect\astroncite{{White}}{2006}]{White06}
{White}, M.: 2006,
\newblock {\em New Astronomy Review} {\bf 50}, 938

\bibitem[\protect\astroncite{{Wu} and {Wang}}{1986}]{Wu86}
{Wu}, Y. and {Wang}, Z.: 1986,
\newblock {\em Physical Review Letters} {\bf 57}, 1978

\bibitem[\protect\astroncite{{Yoo} and {Scherrer}}{2003}]{YS03}
{Yoo}, J.~J. and {Scherrer}, R.~J.: 2003,
\newblock {\em Phys.Rev.D} {\bf 67(4)}, 043517

\bibitem[\protect\astroncite{{Youm}}{2001a}]{Youm2001a}
{Youm}, D.: 2001a,
\newblock {\em Phys.Rev.D} {\bf 63}, 125011

\bibitem[\protect\astroncite{{Youm}}{2001b}]{Youm2001b}
{Youm}, D.: 2001b,
\newblock {\em Phys.Rev.D} {\bf 64}, 085011

\end{thebibliography}
\bibliographystyle{astron}

\end{document}